\documentstyle[11pt,amsmath,amssymb,color,slashed,cite]{article}

\newcommand{\pb}{{\bf p}}
\newcommand{\0}{{\bf 0}}
\newcommand{\xib}{{\boldsymbol{\xi}}}
\newcommand{\ab}{{\bf a}}
\newcommand{\bb}{{\bf b}}

\renewcommand{\d}{{\mathrm{d}}}
\renewcommand{\i}{{\mathrm{i}}}
\newcommand{\Tr}{{\mathrm{Tr}}}
\newcommand{\WW}{{\mathrm{WW}}}

\newcommand{\Wc}{{\mathcal{W}}}

\renewcommand{\>}{\rangle}

\newcommand{\Sb}{{\bf S}}
\newcommand{\Fc}{{\mathcal{F}}}

\begin{document}

%
% Title, Authors, Affiliations 
% ===================
%
\title{\Large\bf 
 Lorentz invariance relations between parton distributions \\
 and  the Wandzura-Wilczek approximation}

\author{A.~Metz$^{1}$, P.~Schweitzer$^{2}$, and T.~Teckentrup$^{3}$
 \\[0.3cm]
{\normalsize\it $^1$Department of Physics, Barton Hall,} %\\
{\normalsize\it Temple University, Philadelphia, PA 19122-6082, USA} \\[0.1cm]
{\normalsize\it $^2$Department of Physics,} %\\
{\normalsize\it University of Connecticut, Storrs, CT 06269, USA} \\[0.1cm]
{\normalsize\it $^3$Institut f{\"u}r Theoretische Physik II,} %\\
{\normalsize\it Ruhr-Universit{\"a}t Bochum, 44780 Bochum, Germany}}

\maketitle

% headline
%\thispagestyle{fancy}
%\fancyhead[R]{\tt \jobname}

%
% Abstract
% ======
%
\begin{abstract}
\noindent
The violation of the so-called Lorentz invariance relations between parton distribution 
functions is considered in a model independent way. 
It is shown that these relations are not violated in a generalized Wandzura-Wilczek 
approximation, indicating that numerically their violation may be small. 
\end{abstract}

%
% 1. Section: Introduction
% ==================
%
\section{Introduction}
\noindent
Parton distribution functions (PDFs) which are of higher twist and/or which are 
transverse momentum dependent ($\pb_T$-dependent) contain important information on 
the partonic structure of the nucleon being complementary to that encoded in the usual 
twist-2 distributions. 
These PDFs also become more important because of the increasing accuracy of recent and 
planned high energy scattering experiments. 
The forward twist-3 PDFs are accessible through certain spin asymmetries in polarized 
inclusive deep inelastic lepton nucleon scattering (DIS) and Drell-Yan 
processes (integrated upon the transverse momentum of the dilepton 
pair)~\cite{Jaffe:1991kp,Tangerman:1994bb,Adams:1994id,Abe:1998wq,Anthony:2002hy,Amarian:2003jy,Zheng:2004ce,Kramer:2005qe}. 
On the other hand the $\pb_T$-dependent PDFs typically give rise to spin and azimuthal 
asymmetries in, for instance, semi-inclusive 
DIS~\cite{Mulders:1995dh,Boer:1997nt,Barone:2001sp,Bacchetta:2006tn,D'Alesio:2007jt} 
and Drell-Yan~\cite{Ralston:1979ys,Tangerman:1994eh,Boer:1999mm,Arnold:2008kf}, 
and significant effort has already been devoted to measure such 
observables~\cite{Airapetian:1999tv,Avakian:2003pk,Airapetian:2004tw,Alexakhin:2005iw,Diefenthaler:2005gx,Ageev:2006da,Diefenthaler:2007rj,Alekseev:2008dn,Levorato:2008tv,Falciano:1986wk,Guanziroli:1987rp,Conway:1989fs,Zhu:2006gx}.

Several relations between (forward) twist-3 and (moments of) $\pb_T$-dependent PDFs 
have been proposed in the literature~\cite{Tangerman:1994bb,Mulders:1995dh,Boer:1997nt}. 
The derivation of these relations is based upon the general, Lorentz invariant 
decomposition of the fully unintegrated correlator of two quark-fields, where the 
fields are located at arbitrary space-time positions. 
These so-called Lorentz invariance relations (LIRs) impose important constraints on the 
PDFs which may allow one to eliminate unknown PDFs in favor of the known ones 
whenever applicable. 
However, in Refs.~\cite{Kundu:2001pk,Schlegel:2004rg} it was demonstrated by an 
explicit model calculation that two specific LIRs are actually violated.
In~\cite{Goeke:2003az} it was shown in a model independent way that the violation can 
be traced back to the path-ordered exponential in the unintegrated quark-quark 
correlator.
For completeness we will repeat the argument below.

In the present letter we address the question to what extent the LIRs may be violated
{\em numerically}.
To this end it is outlined in a model independent way that the LIRs are actually not 
violated in a generalized Wandzura-Wilczek approximation where one systematically 
neglects certain quark-gluon-quark correlations as well as current quark mass terms. 
This result indicates that the numerical violation of the LIRs may be rather 
small and could even be neglected in special cases. 

%
% 2. Section: LIRs and their violation
% ===========================
%
\section{Lorentz invariance relations and their violation}
\noindent
In order to discuss the LIRs and their violation we start with the fully unintegrated 
quark-quark correlation function of a spin-$\frac{1}{2}$ hadron defined 
by\footnote{Recent work on how the unintegrated quark-quark correlator may enter
observables can be found in~\cite{Collins:2007ph,Rogers:2008jk}.}
\begin{equation} \label{eq:corr}
\Phi_{ij}(P,p,S|n_-) = \int \frac{\d^4\xi}{(2\pi)^4} \, e^{\i p \cdot \xi} \, 
\<P,S \, | \, \bar{\psi}_j(0) \, \Wc (0,\xi|n_-) \, \psi_i(\xi) \, | \, P,S\> \, .
\end{equation}
The target state is characterized by its four-momentum $P = P^+n_+ + (M^2/2P^+)n_-$ 
and the covariant spin vector $S$ ($P^2 = M^2, \; S^2=-1, \; P\cdot S =0$), with 
the two light-like vectors $n_+$ and $n_-$ satisfying $n^2_+ = n^2_- = 0$ and 
$n_+ \cdot n_- = 1$. 
The momentum of the quark is denoted by $p$. 
The Wilson line $\Wc (0,\xi|n_-)$ ensures color gauge invariance of the correlator, 
where the specific path of the gauge link will be given below. 
The knowledge of the correlator in Eq.~(\ref{eq:corr}) is particularly useful for 
obtaining the general form of the $\pb_T$-dependent correlator 
$\Phi(x,\pb_T,S)$, which appears in the QCD-description of hard scattering processes 
like transverse momentum dependent semi-inclusive DIS and the Drell-Yan reaction. 
The connection between both objects is given by the relation
\begin{equation} \label{eq:corr_int}
\Phi(x,\pb_T,S) = \int \d p^- \, \Phi(P,p,S|n_-) \, ,
\end{equation}
with $x$ defining the plus-momentum of the quark via $p^+ =x P^+$.
The Wilson line in Eq.~(\ref{eq:corr}) can be fixed according to
\begin{equation} \label{eq:wilson}
\Wc(0,\xi|n_-) = [0,0,\0_T;0,\infty,\0_T] 
 \times [0,\infty,\0_T;\xi^+,\infty,\xib_T] 
 \times [\xi^+,\infty,\xib_T;\xi^+,\xi^-,\xib_T] \, ,
\end{equation}
where $[a^+,a^-,\ab_T;b^+,b^-,\bb_T]$ denotes a gauge link connecting the points 
$a^{\mu}=(a^+,a^-,\ab_T)$ and $b^{\mu}=(b^+,b^-,\bb_T)$ along a straight line. 
It is important to note that the Wilson contour in Eq.~(\ref{eq:wilson}) not only 
depends on the coordinates of the initial and final points but also on the 
light-cone direction~$n_-$. 
The path is chosen such that, upon integration over the minus-momentum of the quark, 
it leads to a proper definition of the $\pb_T$-dependent correlator 
in~(\ref{eq:corr_int})~\cite{Ji:2002aa,Belitsky:2002sm,Boer:2003cm}. 
The choice of the contour depends on the process under 
consideration~\cite{Collins:2002kn}. 
Here we restrict ourselves to the case of semi-inclusive DIS, but all arguments 
hold as well for other processes like Drell-Yan.
It is also worthwhile to mention that Wilson lines that are near the light-cone 
rather than exactly light-like (as those in (\ref{eq:wilson})) are generally 
more appropriate in connection with unintegrated parton correlation functions.
(More details on this important issue can be found in the recent 
work~\cite{Collins:2008ht} and references therein.)
Our general reasoning here remains valid if one uses a near light-cone direction 
instead of $n_{-}$.

The general structure of the correlator in~(\ref{eq:corr}) was derived 
in~\cite{Goeke:2005hb}\footnote{For a spin-0 hadron see  
Refs.~\cite{Goeke:2003az,Bacchetta:2004zf}.}, and the result is also given in 
Eq.~(\ref{eq:app_corr}) in App.~\ref{App-A}.
One ends up with 32 matrix structures multiplied by scalar functions that were
denoted by $A_i$ and $B_i$ in~\cite{Goeke:2005hb}. 
In turn, the $\pb_T$-dependent PDFs can be defined through Dirac traces of the 
$\pb_T$-dependent correlator in~(\ref{eq:corr_int}) given 
by~\cite{Mulders:1995dh,Goeke:2005hb,Bacchetta:2006tn}
\begin{equation} \label{eq:trace}
\Phi^{[\Gamma]}(x,\pb_T,S)\equiv \frac{1}{2}\Tr\big(\Phi(x,\pb_T,S)\,\Gamma \big) \, .
\end{equation}
The results containing all the twist-2 and twist-3 PDFs are repeated
in App.~\ref{App-A}, Eqs.~(\ref{eq:app_corr_tr1})--(\ref{eq:app_corr_tr9}).
On the basis of the relation~(\ref{eq:corr_int}) one can now express the 
$\pb_T$-dependent PDFs through $p^-$-integrals upon the scalar functions $A_i$
and $B_i$. 
The results are listed in App.~\ref{App-A}, 
Eqs.~(\ref{eq:app_f_1})--(\ref{eq:app_h_T}).

In total there exist 32 $\pb_T$-dependent PDFs which exactly agrees with the 
number of independent amplitudes $A_i$ and $B_i$. 
If one neglects the dependence on the light-cone vector $n_{-}$, which is 
induced by the Wilson line, the correlator~(\ref{eq:corr}) merely consists of 
12 matrix structures --- those which are multiplied by the functions $A_i$.
In that case the number of $\pb_T$-dependent PDFs is larger than the number of 
the amplitudes $A_i$. 
This feature, in particular, gives rise to LIRs between certain 
$\pb_T$-integrated PDFs and (moments of) $\pb_T$-dependent 
PDFs~\cite{Tangerman:1994bb,Mulders:1995dh,Boer:1997nt}. 
Here we list the most important four LIRs on which we focus in this work:
\begin{eqnarray} \label{eq:LIR1}
g_T(x) \; &\stackrel{\text{LIR}}{=}& \; 
  g_1(x) + \frac{\d}{\d x} g^{(1)}_{1T}(x) \, , \\
\label{eq:LIR2}
h_L(x) \; &\stackrel{\text{LIR}}{=}& \; 
  h_1(x) - \frac{\d}{\d x} h^{\perp(1)}_{1L}(x) \, , \\
\label{eq:LIR3}
f_T(x) \; &\stackrel{\text{LIR}}{=}& \; 
  - \frac{\d}{\d x} f^{\perp(1)}_{1T}(x) \, , \\
\label{eq:LIR4}
h(x) \; &\stackrel{\text{LIR}}{=}& \; 
  - \frac{\d}{\d x} h^{\perp(1)}_1(x) \, ,
\end{eqnarray}
with
\begin{equation} \label{eq:moments}
g^{(1)}_{1T}(x) \; = \; \int \d^2 \pb_T \, \frac{\pb^2_T}{2M^2} \, g_{1T}(x,\pb^2_T) \,, 
\; \text{etc.}
\end{equation}
specifying certain $\pb_T$-moments of the PDFs~\cite{Mulders:1995dh}. 
All PDFs on the \textit{lhs} of Eqs.~(\ref{eq:LIR1})--(\ref{eq:LIR4}) are twist-3
functions, while those on the \textit{rhs} are  not suppressed in observables. 
The PDFs in Eqs.~(\ref{eq:LIR1}) and (\ref{eq:LIR2}) are time-reversal even (T-even), 
and the ones in Eqs.~(\ref{eq:LIR3}) and (\ref{eq:LIR4}) are T-odd.
The derivation of the LIR (\ref{eq:LIR4}) is sketched in App.~\ref{App-A} in 
Eqs.~(\ref{eq:app_h_pi})--(\ref {eq:app_d_h^perp1_1_pi}).

If, however, one takes into account the Wilson line ($n_{-}$-dependence) in the 
correlator~(\ref{eq:corr}) the LIRs~(\ref{eq:LIR1})--(\ref{eq:LIR4}) are no longer 
fulfilled, i.e., they are spoiled by the presence of the amplitudes $B_i$.
The reasoning shows in a model independent way that in the context of a gauge 
theory, for which the Wilson line is mandatory, the LIRs are violated.
This is the general message of Ref.~\cite{Goeke:2003az}.

%
% 3. Section: LIRs in a generalized WW approximation
% =======================================
%
\section{Lorentz invariance relations in a generalized Wandzura-Wilczek approximation}
\noindent
Knowing that the LIRs are violated it is now natural to ask to what extent they 
are violated {\em numerically}.
If one gets an indication that the violation of the LIRs should be small, these 
relations can still serve as a useful tool --- at least for qualitative studies 
of the partonic nucleon structure.
In what follows we consider the LIRs and their violation in a model independent 
way using a special approximation.
We start with the discussion of the LIRs~(\ref{eq:LIR1}) and (\ref{eq:LIR2}) 
which contain T-even PDFs.

First we recall the following relations between $\pb_T$-integrated T-even 
PDFs~\cite{Wandzura:1977qf,Jaffe:1991kp}:
\begin{eqnarray} \label{eq:coll_pdf1}
g_T(x) \; &=& \; \int^1_x \frac{\d y}{y} g_1(y) + \tilde{g}'_T(x) \, , \\
\label{eq:coll_pdf2}
h_L(x) \; &=& \; 2x \int^1_x \frac{\d y}{y^2} h_1(y) + \tilde{h}'_L(x) \, ,
\end{eqnarray}
where $\tilde{g}'_T(x)$ and $\tilde{h}'_L(x)$ denote (purely interaction dependent)
quark-gluon-quark correlations and terms proportional to current quark masses. 
An explicit representation of these terms can be found, e.g., in~\cite{Belitsky:1997ay} 
and partly also in~\cite{Accardi:2009nv}.
Eqs.~(\ref{eq:coll_pdf1}) and (\ref{eq:coll_pdf2}) isolate ``pure twist-3 terms'' 
in the PDFs $g_T(x)$ and $h_L(x)$. 
Here the underlying ``working definition'' of twist~\cite{Jaffe:1996zw} 
(a PDF is of ``twist $t$'' if its contribution to the cross section is suppressed, 
in addition to kinematic factors, by $1/Q^{t-2}$ with $Q$ denoting the hard scale 
of the process) differs from the strict definition of twist (mass dimension of 
the operator minus its spin).

The remarkable experimental observation is that $\tilde{g}'_T(x)$ is consistent 
with zero within the error 
bars~\cite{Adams:1994id,Abe:1998wq,Anthony:2002hy,Amarian:2003jy,Zheng:2004ce,Kramer:2005qe} 
and to good accuracy one has
\begin{equation} \label{eq:WW1}
g_T(x) \; \stackrel{\WW}{\approx} \; \int^1_x \frac{\d y}{y} \; g_1(y) \,,
\end{equation}
which is the Wandzura-Wilczek (WW) approximation. 
Lattice QCD~\cite{Gockeler:2000ja,Gockeler:2005vw} and the instanton model of the 
QCD vacuum~\cite{Balla:1997hf} support this observation. 
(Further discussions of the WW approximation in related and other contexts can be 
found in Refs.~\cite{Ball:1996tb,Blumlein:1996vs,Blumlein:1998nv,Kivel:2000rb,Radyushkin:2000ap,Anikin:2001ge,Teryaev:1995um}.)
Interestingly, the latter predicts also $\tilde{h}'_L(x)$ to be 
small~\cite{Dressler:1999hc} such that
\begin{equation} \label{eq:WW2}
h_L(x) \; \stackrel{\WW}{\approx} \; 2x \int^1_x \frac{\d y}{y^2} \; h_1(y) \, .
\end{equation}
An experimental test of this approximate relation does not yet exist.
(In this context see also the recent theoretical study in Ref.~\cite{Koike:2008du}.)

Now it is possible to show that the LIRs in Eqs.~(\ref{eq:LIR1}) and (\ref{eq:LIR2}) 
are not violated if one generalizes the WW approximation.
For this purpose we consider the following exact 
relations~\cite{Mulders:1995dh,Bacchetta:2006tn} originating from the QCD 
equations of motion (EOM):
\begin{eqnarray} \label{eq:EOM1}
g^{(1)}_{1T}(x) \; &\stackrel{\text{EOM}}{=}& 
\; x\;(g_T(x)-\tilde{g}_T(x)) - \frac{m}{M}\;h_1(x) \, , \\
\label{eq:EOM2}
h^{\perp(1)}_{1L}(x) \; &\stackrel{\text{EOM}}{=}& 
\;-\frac{x}{2} \;(h_L(x)-\tilde{h}_L(x)) + \frac{m}{2M}\;g_{1L}(x) \, ,
\end{eqnarray}
with the (1)-moments of the PDFs defined in Eq.~(\ref{eq:moments}). 
The functions $\tilde{g}_T(x)$ and $\tilde{h}_L(x)$ denote twist-3 
quark-gluon-quark correlations. 
In lightcone gauge these objects, like $\tilde{g}'_T$ and $\tilde{h}'_L$, 
represent matrix elements of the type 
$\langle \, | \bar{\Psi} A_T \Psi | \,\rangle$.
One therefore can assume that these functions are small as well, although the
explicit form of $\tilde{g}_T \; (\tilde{h}_L)$ differs from the one of 
$\tilde{g}'_T \; (\tilde{h}'_L)$. 
In the following we denote as ``WW-type approximation'' the neglect of the 
tilde-functions (and quark mass terms) in the EOM-relations.\footnote{Note 
that in Ref.~\cite{Bacchetta:2006tn} for instance this approximation  
(for brevity) was just called ``WW approximation'' because, like 
Eqs.~(\ref{eq:WW1}) and~(\ref{eq:WW2}), it also corresponds to neglecting 
purely interaction dependent terms.}
Below we will also address the phenomenological justification of the 
WW-type approximation.

In order to proceed we introduce a measure $\Delta_{g}(x)$ and a measure 
$\Delta_{h}(x)$ for the violation of the LIRs (see for instance 
Refs.~\cite{Belitsky:1997ay,Accardi:2009nv} for explicit forms of these terms) 
according to
\begin{eqnarray} \label{eq:LIR1_delta}
g_T(x) \; &=& \; g_1(x) + \frac{\d}{\d x} g^{(1)}_{1T}(x) + \Delta_{g}(x) \, , \\
\label{eq:LIR2_delta}
h_L(x) \; &=& \; h_1(x) - \frac{\d}{\d x} h^{\perp(1)}_{1L}(x) + \Delta_{h}(x) \, .
\end{eqnarray}
If one substitutes the (1)-moments $g^{(1)}_{1T}$ and $h^{\perp(1)}_{1L}$
from~(\ref{eq:EOM1}) and (\ref{eq:EOM2}) in Eqs.~(\ref{eq:LIR1_delta}) and 
(\ref{eq:LIR2_delta}), and uses both the WW approximation and the WW-type
approximation one finds
\begin{eqnarray} \label{eq:delta1}
\Delta_{g}(x) \; &\stackrel{\mathrm{WW,WW-type}}{\approx}& \; 
-g_1(x) - x \; \frac{\d}{\d x} \int^1_x \frac{\d y}{y} \; g_1(y) \;\;\; = \; 0 \, , \\
\label{eq:delta2}
\Delta_{h}(x) \; &\stackrel{\mathrm{WW,WW-type}}{\approx}& \; 
-h_1(x) - x^2 \; \frac{\d}{\d x} \int^1_x \frac{\d y}{y^2} \; h_1(y) \; = \; 0 \, .
\end{eqnarray}
Eqs.~(\ref{eq:delta1}) and (\ref{eq:delta2}) show that the LIRs~(\ref{eq:LIR1}) and 
(\ref{eq:LIR2}) are not violated if a generalized WW approximation is applied.
This result is not entirely surprising keeping in mind that the violation of the
LIRs is related to the amplitudes $B_i$ which are associated with the gauge link
of the fully unintegrated correlator in~(\ref{eq:corr}).
Therefore, the $B_i$'s are necessarily related to quark-gluon-quark correlations 
which are neglected in the (generalized) WW approximation.
This is also in line with the fact that in relativistic nucleon models without 
gluonic degrees of freedom the LIRs are typically fulfilled 
(see, e.g., Refs.~\cite{Jakob:1997wg,Efremov:2009ze}).

We also point out that in the generalized WW approximation instead of having 
the three functions $g_1$, $g_T$, and $g_{1T}^{(1)}$, there is only one 
independent PDF.
The same applies to the $h$-functions.
In particular, one can immediately write
\begin{eqnarray} \label{eq:mom_g}
g_{1T}^{(1)}(x) \; &\stackrel{\mathrm{WW,WW-type}}{\approx}& \; 
x \int^1_x \frac{\d y}{y} g_1(y) \,,
\\ \label{eq:mom_h}
h_{1L}^{\perp(1)}(x) \; &\stackrel{\mathrm{WW,WW-type}}{\approx}& \; 
- x^2 \int^1_x \frac{\d y}{y^2} h_1(y) \,.
\end{eqnarray}
Phenomenological work on the basis of those relations was, for instance, carried 
out in~\cite{Kotzinian:1995cz,Kotzinian:2006dw,Avakian:2007mv}. 
In Ref.~\cite{Avakian:2007mv}, Eq.~(\ref{eq:mom_h}) was used in order to 
describe data for a certain longitudinal single spin asymmetry in semi-inclusive 
DIS. 
This investigation shows that the approximation~(\ref{eq:mom_h}) is not excluded, 
although more precise data would be helpful for having a stronger test.
Further support for the generalized WW approximation --- and therefore also 
justification for the WW-type approximation --- comes from 
studies~\cite{Anselmino:2005nn} of the azimuthal $\cos{\Phi}$-dependence of the 
unpolarized cross section for semi-inclusive DIS on the basis of the Cahn 
effect~\cite{Cahn:1978se,Cahn:1989yf}.
This approach, though neglecting tilde-functions from the QCD equations of 
motion (and the Boer-Mulders effect, see Ref.~\cite{Bacchetta:2006tn} 
for more details), provides a quite satisfactory description of the available 
data~\cite{Anselmino:2005nn}. 

One has to keep in mind that neglecting {\sl several} tilde-functions as done in 
the last example, even if each by itself is small, may naturally lead to a poor 
approximation.
Two examples from semi-inclusive DIS at subleading twist illustrate this.
In the longitudinal target spin asymmetry proportional to $\sin\phi$ of the 
produced hadron, the `overeager' use of the WW-approximation would imply that the 
observable is due to the Collins effect only, but data \cite{Airapetian:1999tv} 
do not follow the flavor pattern observed for the pure Collins effect
\cite{Airapetian:2004tw,Diefenthaler:2005gx,Diefenthaler:2007rj}.
The use of the WW-approximation is also not recommended in the case of the beam 
spin asymmetry, in which the {\sl entire} effect is due to tilde-functions. 
The data on this small but clearly non-zero asymmetry \cite{Avakian:2003pk} provide 
direct access to tilde-functions and hence quark-gluon-quark correlations.
In both cases the asymmetries are small, which means that there does not need to 
be a general conflict with the WW-approximation.
\\[0.5cm]
\noindent
In case of the LIRs~(\ref{eq:LIR3}) and (\ref{eq:LIR4}) which contain T-odd PDFs 
the situation is slightly different and in principle even simpler. 
Due to time-reversal invariance the $\pb_T$-integrated T-odd PDFs $f_T(x)$ and 
$h(x)$ vanish~\cite{Goeke:2005hb,Bacchetta:2006tn}, 
\begin{eqnarray} \label{eq:T_odd1}
f_T(x) \; &=& \; \int \d^2 \pb_T \; f_T(x,\pb^2_T) \; = \; 0 \, , \\
\label{eq:T_odd2}
h(x) \; &=& \; \int \d^2 \pb_T \; h(x,\pb^2_T) \; = \; 0 \, ,
\end{eqnarray}
which implies, considering the LIRs in Eqs.~(\ref{eq:LIR3}) and (\ref{eq:LIR4}),
that 
\begin{eqnarray} \label{eq:LIR3_odd}
\frac{\d}{\d x} \; f^{\perp(1)}_{1T}(x) \; &\stackrel{\text{LIR}}{=}& \; 0 \, , \\
\label{eq:LIR4_odd}
\frac{\d}{\d x} \; h^{\perp(1)}_1(x) \; &\stackrel{\text{LIR}}{=}& \; 0 \, .
\end{eqnarray}
This means that $f^{\perp(1)}_{1T}$ and $h^{\perp(1)}_1$ are constants.
In fact, since these moments have to vanish for $x=1$, one can conclude 
that they should vanish for the entire $x$-range. 
So far we did not use any approximation and only assumed that the 
LIRs~(\ref{eq:LIR3}) and (\ref{eq:LIR4}) are not violated. 
Now let us explore the EOMs \cite{Bacchetta:2006tn} which, keeping in mind 
(\ref{eq:T_odd1}) and (\ref{eq:T_odd2}), imply 
(see also Refs.~\cite{Boer:2003cm,Ma:2003ut,Ji:2006ub})
\begin{eqnarray} \label{eq:EOM3}
f^{\perp(1)}_{1T}(x) \; &\stackrel{\text{EOM}}{=}& \; x \; \tilde{f}_T(x) \, , \\
\label{eq:EOM4}
h^{\perp(1)}_1(x) \; &\stackrel{\text{EOM}}{=}& \; \frac{x}{2} \; \tilde{h}(x) \, .
\end{eqnarray}
In the WW-type approximation the tilde-functions are set to zero.
It then follows directly from Eqs.~(\ref{eq:EOM3}) and (\ref{eq:EOM4}) that
$f^{\perp(1)}_{1T}$ and $h^{\perp(1)}_1$ are zero (see also~\cite{Bacchetta:2006tn}).
This is consistent with the results following from the LIRs~(\ref{eq:LIR3_odd}) 
and (\ref{eq:LIR4_odd}).
So the LIRs~(\ref{eq:LIR3}) and (\ref{eq:LIR4}) are also not violated in the 
WW-type approximation.

Also for the T-odd functions we already have some phenomenological input on the
status of the WW-type approximation.
Since a nonzero asymmetry, typically attributed to the Sivers effect, was found 
in the HERMES experiment~\cite{Airapetian:2004tw,Diefenthaler:2005gx,Diefenthaler:2007rj}, 
in the case of T-odd PDFs this approximation seems to be violated. 
On the other hand the observed effect is not very large (of the order of 
few percent), and one should not expect WW-type approximations to work to a 
much better accuracy than that.
Moreover, the Sivers effect studied at COMPASS is compatible with zero both 
for a deuteron as well as a proton 
target~\cite{Alexakhin:2005iw,Ageev:2006da,Alekseev:2008dn,Levorato:2008tv}.
Therefore, the current experimental situation is not in conflict with a rather 
small $\tilde{f}_T$ in Eq.~(\ref{eq:EOM3}).

%
% 4. Section: Summary
% ================
%
\section{Summary}
\noindent
We have studied LIRs between parton distributions, known to be violated
in general, with the aim to understand how strong this violation might be.
It was found that LIRs are {\em satisfied} in a generalized WW approximation
in which one systematically neglects certain quark-gluon-quark correlations
as well as quark mass terms.
That would mean that LIRs could provide useful approximations for unknown 
PDFs whenever applicable.
Our approximation goes beyond the successful ``standard WW approximation'' 
quoted in Eqs.~(\ref{eq:WW1}) and (\ref{eq:WW2}).
In particular, we also neglected purely interaction dependent terms which 
show up in relations originating from the QCD equations of motion 
(see also Ref.~\cite{Bacchetta:2006tn}).
We argued that there exists experimental evidence for the validity of 
the generalized WW approximation.
On the other hand more (precise) data and tests are needed before a final
conclusion can be reached.
Only forthcoming data analyses and experiments at COMPASS, HERMES, and 
Jefferson Lab can ultimately reveal to what extent the generalized WW 
approximation (and the LIRs) provide useful approximations.
Eventually, it is likely that the quality of the approximation depends on 
the particular case (function) under consideration. 
\\[0.5cm]
%
% Acknowledgments
% ===============
%
\noindent
{\bf Acknowledgments:} 
We are grateful to A.~Accardi, A.~Bacchetta, J.~Bl\"umlein, S.~Mei{\ss}ner,
M.~V.~Polyakov, M.~Schlegel, and O.~V.~Teryaev for discussions.
The work is partially supported by the Verbundforschung 
``Hadronen und Kerne'' of the BMBF. 
A.M. acknowledges the support of the NSF under Grant No. PHY-0855501.
P.S. is supported by DOE contract No. DE-AC05-06OR23177, under
which Jefferson Science Associates, LLC operates Jefferson Lab.
T. T. is supported by the Cusanuswerk.
\\[0.5cm]
{\bf Note added:}
After completion of our work the manuscript~\cite{Accardi:2009nv} appeared where,
on the basis of the present data for the DIS structure function $g_2$, a violation 
of the WW-relation in Eq.~(\ref{eq:WW1}) of the order $15-40\%$ has been reported.
However, the authors of~\cite{Accardi:2009nv} also point out that more data are 
needed to ultimately settle the situation.
In any case, if generalized WW-relations were valid within a similar accuracy, 
they would constitute helpful tools at the present stage for phenomenological studies 
of first data.

%
% Appendix
% ========
%
\appendix
\section{Quark-quark correlators and the Lorentz invariance relations}
\label{App-A}

\noindent
For completeness we give here the general structure of the fully unintegrated
quark-quark correlator in Eq.~(\ref{eq:corr}), the $\pb_T$-dependent 
quark-quark correlator (expressed in terms of $\pb_T$-dependent PDFs) in 
Eq.~(\ref{eq:corr_int}), the relations between PDFs and the amplitudes 
$A_i$, $B_i$ which parameterize the correlator in~(\ref{eq:corr}), as well 
as a brief account on how to derive the LIRs in the case that the amplitudes 
$B_i$ are absent (like in a non-gauge theory).

The fully unintegrated quark-quark correlator~(\ref{eq:corr}) can be decomposed
according to~\cite{Goeke:2005hb}
\begin{eqnarray}
\label{eq:app_corr}
\Phi(P,p,S|n_-) \; &= \;& MA_1 + \slashed{P} A_2 + \slashed{p}A_3 + \frac{\i}{2M} \; [\slashed{P},\slashed{p}] \; A_4 + \i (p \cdot S) \gamma_5 \; A_5 + M\slashed{S} \gamma_5 \; A_6\\
&&+ \frac{p \cdot S}{M} \slashed{P} \gamma_5 \; A_7 + \frac{p \cdot S}{M} \slashed{p} \gamma_5 \; A_8 + \frac{[\slashed{P},\slashed{S}]}{2} \gamma_5 \; A_9 + \frac{[\slashed{p},\slashed{S}]}{2} \gamma_5 \; A_{10} \nonumber \\
&&+ \frac{p \cdot S}{2M^2} [\slashed{P},\slashed{p}] \gamma_5 \; A_{11} + \frac{1}{M} \varepsilon^{\mu\nu\rho\sigma} \gamma_{\mu} P_{\nu} p_{\rho} S_{\sigma} \; A_{12} \nonumber \\
&&+ \frac{M^2}{P\cdot n_-} \slashed{n}_- \; B_1 + \frac{\i M}{2 P \cdot n_-} [\slashed{P},\slashed{n}_-] \; B_2 + \frac{\i M}{2 P \cdot n_-} [\slashed{p},\slashed{n}_-] \; B_3 \nonumber \\
&&+ \frac{1}{P \cdot n_-} \varepsilon^{\mu\nu\rho\sigma} \gamma_{\mu} \gamma_5 P_{\nu} p_{\rho} n_{-\sigma} \; B_4 \nonumber \\
&&+ \frac{1}{P \cdot n_-} \varepsilon^{\mu\nu\rho\sigma} P_{\mu} p_{\nu} n_{-\rho} S_{\sigma} \; B_5 + \frac{\i M^2}{P \cdot n_-} (n_- \cdot S) \gamma_5 \; B_6 \nonumber \\
&&+ \frac{M}{P \cdot n_-} \varepsilon^{\mu\nu\rho\sigma} \gamma_{\mu} P_{\nu} n_{-\rho} S_{\sigma} \; B_7 + \frac{M}{P \cdot n_-} \varepsilon^{\mu\nu\rho\sigma} \gamma_{\mu} p_{\nu} n_{-\rho} S_{\sigma} \; B_8 \nonumber \\
&&+ \frac{p \cdot S}{M(P \cdot n_-)} \varepsilon^{\mu\nu\rho\sigma} \gamma_{\mu} P_{\nu} p_{\rho} n_{-\sigma} \; B_9 + \frac{M(n_- \cdot S)}{(P \cdot n_-)^2} \varepsilon^{\mu\nu\rho\sigma} \gamma_{\mu} P_{\nu} p_{\rho} n_{-\sigma} \; B_{10} \nonumber \\
&&+ \frac{M}{P \cdot n_-} (n_- \cdot S) \slashed{P} \gamma_5 \; B_{11} + \frac{M}{P \cdot n_-} (n_- \cdot S) \slashed{p} \gamma_5 \; B_{12} \nonumber \\
&&+ \frac{M}{P \cdot n_-} (p \cdot S) \slashed{n}_- \gamma_5 \; B_{13} + \frac{M^3}{(P \cdot n_-)^2} (n_- \cdot S) \slashed{n}_- \gamma_5 \; B_{14} \nonumber \\
&&+ \frac{M^2}{2P \cdot n_-} [\slashed{n}_-,\slashed{S}] \gamma_5 \; B_{15} + \frac{p \cdot S}{2P \cdot n_-} [\slashed{P},\slashed{n}_-] \gamma_5 \; B_{16} + \frac{p \cdot S}{2P \cdot n_-} [\slashed{p},\slashed{n}_-] \gamma_5 \; B_{17} \nonumber \\
&&+ \frac{n_- \cdot S}{2P \cdot n_-} [\slashed{P},\slashed{p}] \gamma_5 B_{18} + \frac{M^2(n_- \cdot S)}{2(P \cdot n_-)^2} [\slashed{P},\slashed{n}_-] \gamma_5 B_{19} + \frac{M^2(n_- \cdot S)}{2(P \cdot n_-)^2} [\slashed{p},\slashed{n}_-] \gamma_5 B_{20} \,, \nonumber
\end{eqnarray}
where the convention $\varepsilon^{0123} = 1$ is understood.
The scalar amplitudes depend on the available kinematical invariants.
Note that all the $B_i$'s are associated with matrix structures containing the light-like
vector $n_{-}$.

The $\pb_T$-dependent correlator in Eq.~(\ref{eq:corr_int}) can be specified by all
possible Dirac traces $\Phi^{[\Gamma]}$ defined in Eq.~(\ref{eq:trace}), which in 
turn are parameterized through $\pb_T$-dependent PDFs.
A list of all traces was given in Refs.~\cite{Goeke:2005hb,Bacchetta:2006tn}.
Limiting oneself to twist-2 and twist-3 effects one has 
(in the conventions of~\cite{Bacchetta:2006tn}):
\begin{eqnarray}
\label{eq:app_corr_tr1}
\Phi^{[\gamma^+]}(x,\pb_T,S) \; &=& \; f_1(x,\pb^2_T) - \frac{\varepsilon^{ij}_T \, p_{Ti} \, S_{Tj}}{M} \; f^{\perp}_{1T}(x,\pb^2_T) \, , \\
\label{eq:app_corr_tr2}
\Phi^{[\gamma^+ \gamma_5]}(x,\pb_T,S) \; &=& \; \lambda \; g_{1L}(x,\pb^2_T) + \frac{\pb_T \cdot \Sb_T}{M} \; g_{1T}(x,\pb^2_T) \, , \\
\label{eq:app_corr_tr3}
\Phi^{[\i\sigma^{i+}\gamma_5]}(x,\pb_T,S) \; &=& \; S^i_T \; h_1(x,\pb^2_T) + \lambda \; \frac{p^i_T}{M} \; h^{\perp}_{1L}(x,\pb^2_T) \\
&& - \frac{p^i_T \, p^j_T + \frac{1}{2} \, \pb^2_T \, g^{ij}_T}{M^2} \; S_{Tj} \; h^{\perp}_{1T}(x,\pb^2_T) - \frac{\varepsilon^{ij}_T \, p_{Tj}}{M} \; h^{\perp}_1(x,\pb^2_T) \, , 
\nonumber \\
\label{eq:app_corr_tr4}
\Phi^{[1]}(x,\pb_T,S) \; &=& \; \frac{M}{P^+} \biggl[ e(x,\pb^2_T) - \frac{\varepsilon^{ij}_T \, p_{Ti} \, S_{Tj}}{M} \; e^{\perp}_T(x,\pb^2_T) \biggr] \, , \\
\label{eq:app_corr_tr5}
\Phi^{[\i \gamma_5]}(x,\pb_T,S) \; &=& \; \frac{M}{P^+} \biggl[ \lambda \; e_L(x,\pb^2_T) + \frac{\pb_T \cdot \Sb_T}{M} \; e_T(x,\pb^2_T) \biggr] \, , 
\end{eqnarray}
\begin{eqnarray}
\label{eq:app_corr_tr6}
\Phi^{[\gamma^i]}(x,\pb_T,S) \; &=& \; \frac{M}{P^+} \Biggl[ -\varepsilon^{ij}_T \; S_{Tj} \; f_T(x,\pb^2_T) - \lambda \; \frac{\varepsilon^{ij}_T \,  p_{Tj}}{M} \; f^{\perp}_L(x,\pb^2_T) \\
&&\quad \quad \;\; - \frac{p^i_T \, p^j_T + \frac{1}{2} \, \pb^2_T \, g^{ij}_T}{M^2} \; \varepsilon^{}_{Tjk} \; S^k_T \; f^{\perp}_T(x,\pb^2_T) + \frac{p^i_T}{M} \; f^{\perp}(x,\pb^2_T) \Biggr] \, , \nonumber \\
\label{eq:app_corr_tr7}
\Phi^{[\gamma^i \gamma_5]}(x,\pb_T,S) \; &=& \; \frac{M}{P^+} \Biggl[ S^i_T \; g_T(x,\pb^2_T) + \lambda \; \frac{p^i_T}{M} \; g^{\perp}_L(x,\pb^2_T) \\
&&\quad \quad \;\; -\frac{p^i_T \, p^j_T + \frac{1}{2} \, \pb^2_T \, g^{ij}_T}{M^2} \; S_{Tj} \; g^{\perp}_T(x,\pb^2_T) - \frac{\varepsilon^{ij}_T \,  p_{Tj}}{M} \; g^{\perp}(x,\pb^2_T) \Biggr] \, , \nonumber \\
\label{eq:app_corr_tr8}
\Phi^{[\i \sigma^{ij} \gamma_5]}(x,\pb_T,S) \; &=& \; \frac{M}{P^+} \Biggl[ -\varepsilon^{ij}_T \; h(x,\pb^2_T) + \frac{S^i_T \, p^j_T - p^i_T \, S^j_T}{M} \; h^{\perp}_T(x,\pb^2_T) \Biggr] \, , \\
\label{eq:app_corr_tr9}
\Phi^{[\i \sigma^{+-} \gamma_5]}(x,\pb_T,S) \; &=& \; \frac{M}{P^+} \Biggl[ \lambda \; h_L(x,\pb^2_T) + \frac{\pb_T \cdot \Sb_T}{M} \; h_T(x,\pb^2_T) \Biggr] \, .
\end{eqnarray}
Here use has been made of the Sudakov decompositions
\begin{eqnarray}
S \; & = & \; \lambda \, \frac{P^+}{M} \, n_{+} 
- \lambda \, \frac{M}{2 P^+} \, n_{-} + S_T \,,
\\ 
p \; & = & \; x P^+ \, n_{+} + p^- \, n_{-} + p_T \,,  
\end{eqnarray}
and of $\varepsilon_{T}^{ij} = \varepsilon^{-+ij}$.

On the basis of the relation in~(\ref{eq:corr_int}) it is straightforward to express
the $\pb_T$-dependent PDFs through the amplitudes $A_i$ and $B_i$.
We again restrict ourselves to the twist-2 and twist-3 case for which we obtain
\newpage
\begin{eqnarray}
\label{eq:app_f_1}
f_1(x,\pb^2_T) \; &=& \; 2P^+ \int \d p^- (A_2 + x A_3) \, , \\
\label{eq:app_f^perp_1T}
f^{\perp}_{1T}(x,\pb^2_T) \; &=& \; 2P^+ \int \d p^- A_{12} \, , \\
\label{eq:app_g_1L}
g_{1L}(x,\pb^2_T) \; &=& \; 2P^+ \int \d p^- \biggl( -A_6-\frac{P \cdot p-M^2x}{M^2}(A_7+xA_8)-B_{11}-xB_{12} \biggr) \, , \\
\label{eq:app_g_1T}
g_{1T}(x,\pb^2_T) \; &=& \; 2P^+ \int \d p^- (A_7+xA_8) \, , \\
\label{eq:app_h_1}
h_1(x,\pb^2_T) \; &=& \; 2P^+ \int \d p^- \biggl( -A_9-xA_{10}+\frac{\pb^2_T}{2M^2} \; A_{11} \biggr) \, , \\
\label{eq:app_h^perp_1L}
h^{\perp}_{1L}(x,\pb^2_T) \; &=& \; 2P^+ \int \d p^- \biggl( A_{10} - \frac{P \cdot p - M^2x}{M^2} \; A_{11} - B_{18} \biggr) \, , \\
\label{eq:app_h^perp_1T}
h^{\perp}_{1T}(x,\pb^2_T) \; &=& \; 2P^+ \int \d p^- A_{11} \, , \\
\label{eq:app_h^perp_1}
h^{\perp}_1(x,\pb^2_T) \; &=& \; 2P^+ \int \d p^- (-A_4) \, , \\
\label{eq:app_e}
e(x,\pb^2_T) \; &=& \; 2P^+ \int \d p^- A_1 \, , \\
\label{eq:app_e^perp_T}
e^{\perp}_T(x,\pb^2_T) \; &=& \; 2P^+ \int \d p^- B_5 \, , \\
\label{eq:app_e_L}
e_L(x,\pb^2_T) \; &=& \; 2P^+ \int \d p^- \Biggl( - \frac{P \cdot p - M^2x}{M^2} \; A_5 - B_6 \Biggr) \, , \\
\label{eq:app_e_T}
e_T(x,\pb^2_T) \; &=& \; 2P^+ \int \d p^- A_5 \, , \\
\label{eq:app_f_T}
f_T(x,\pb^2_T) \; &=& \; 2P^+ \int \d p^- \biggl( -\frac{P \cdot p-M^2x}{M^2} \; A_{12} - B_7 - xB_8 - \frac{\pb^2_T}{2M^2} \; B_9 \biggr) \, , \\
\label{eq:app_f^perp_L}
f^{\perp}_L(x,\pb^2_T) \; &=& \; 2P^+ \int \d p^- \biggl( -A_{12}+B_8+\frac{P \cdot p-M^2x}{M^2} \; B_9 +B_{10} \biggr) \, , \\
\label{eq:app_f^perp_T}
f^{\perp}_T(x,\pb^2_T) \; &=& \; 2P^+ \int \d p^- (-B_9) \, , \\
\label{eq:app_f^perp}
f^{\perp}(x,\pb^2_T) \; &=& \; 2P^+ \int \d p^- A_3 \, , \\
\label{eq:appg_T}
g_T(x,\pb^2_T) \; &=& \; 2P^+ \int \d p^- \biggl( -A_6 + \frac{\pb^2_T}{2M^2} \; A_8 \biggr) \, , \\
\label{eq:app_g^perp_L}
g^{\perp}_L(x,\pb^2_T) \; &=& \; 2P^+ \int \d p^- \biggl( - \frac{P \cdot p-M^2x}{M^2}\; A_8 -B_{12} \biggr) \, , \\
\label{eq:app_g^perp_T}
g^{\perp}_T(x,\pb^2_T) \; &=& \; 2P^+ \int \d p^- A_8 \, , \\
\label{eq:app_g^perp}
g^{\perp}(x,\pb^2_T) \; &=& \; 2P^+ \int \d p^- (-B_4) \, , 
\end{eqnarray}
\begin{eqnarray}
\label{eq:app_h}
h(x,\pb^2_T) \; &=& \; 2P^+ \int \d p^- \biggl( \frac{P \cdot p-M^2x}{M^2} \; A_4+B_2+xB_3 \biggr) \, , \\
\label{eq:app_h^perp_T}
h^{\perp}_T(x,\pb^2_T) \; &=& \; 2P^+ \int \d p^- (-A_{10}) \, , \\
\label{eq:app_h_L}
h_L(x,\pb^2_T) \; &=& \; 2P^+ \int \d p^- \biggl( -A_9-\frac{P \cdot p}{M^2} \; A_{10} + \biggl( \frac{P \cdot p-M^2x}{M^2} \biggr)^2 A_{11} \\
&&\quad \quad \quad \quad \quad \quad -B_{15} + \frac{P \cdot p-M^2x}{M^2} \; B_{16} + \frac{P \cdot p-M^2x}{M^2} \; xB_{17} \nonumber \\
&&\quad \quad \quad \quad \quad \quad + \frac{P \cdot p-M^2x}{M^2} \; B_{18} +B_{19}+xB_{20} \biggr) \, , \nonumber \\
\label{eq:app_h_T}
h_T(x,\pb^2_T) \; &=& \; 2P^+ \int \d p^- \biggl( - \frac{P \cdot p-M^2x}{M^2} \; A_{11}-B_{16}-xB_{17} \biggr) \, .
\end{eqnarray}

In the following we give a brief account on how to derive the LIRs in the case that 
the amplitudes $B_i$ are absent using as an example the LIR in Eq.~(\ref{eq:LIR4}). 
(See also the original references~\cite{Tangerman:1994bb,Mulders:1995dh,Boer:1997nt}.) 
Starting from~(\ref{eq:app_h}) one gets
\begin{eqnarray}
h(x) \; &=& \; 2P^+ \int \d p^- \d^2 \pb_T \; 
\biggl( \frac{P \cdot p -M^2x}{M^2} \; A_4 + B_2 + xB_3 \biggr) \nonumber \\
&=& \; \int \d \sigma \d \tau \d^2 \pb_T \; \delta(\tau-x\sigma+M^2x^2+\pb^2_T) \; \biggl( \frac{\sigma -2M^2x}{2M^2} \; A_4 + B_2 + xB_3 \biggr) \nonumber \\
\label{eq:app_h_pi}
&=& \; \pi \int \d \sigma \d \tau \; \Theta(\tau-x\sigma+M^2x^2) \; \biggl( \frac{\sigma -2M^2x}{2M^2} \; A_4 + B_2 + xB_3 \biggr) \, ,
\end{eqnarray}
where $\sigma=2p\cdot P$, $\tau=p^2$, $A_i=A_i(\sigma,\tau)$ and $B_i=B_i(\sigma,\tau)$. 
Using~(\ref{eq:moments}) and~(\ref{eq:app_h^perp_1}) allows one to write
\begin{eqnarray}
h^{\perp(1)}_1(x) \; &=& \; 2P^+ \int \d p^- \d^2 \pb_T \; \frac{\pb^2_T}{2M^2} \; (-A_4) \nonumber \\
\label{eq:app_h^perp_1_sigma}
&=& \; \int \d \sigma \d \tau \d^2 \pb_T \; \frac{\pb^2_T}{2M^2} \; \delta(\tau-x\sigma+M^2x^2+\pb^2_T) \; (-A_4) \, .
\end{eqnarray}
To solve the integral in~(\ref{eq:app_h^perp_1_sigma}) in the same way as 
in~(\ref{eq:app_h_pi}) one can use the relation~\cite{Tangerman:1994bb}
\begin{eqnarray}
\label{eq:app_rel}
\!\!&& \int \d \sigma \d \tau \d^2 \pb_T \; \frac{\pb^2_T}{2M^2} \; \delta(\tau-x\sigma+M^2x^2+\pb^2_T) \; \Fc(x, \sigma,\tau) \\
\!\!&=& \! -\pi \int^1_x \d y \int \d \sigma \d \tau \; \Theta(\tau - y\sigma + M^2y^2) \; \biggl( \frac{\sigma - 2M^2y}{2M^2} \; \Fc(y,\sigma,\tau) - \frac{\tau - y\sigma + M^2y^2}{2M^2} \; \frac{\partial\Fc}{\partial y}(y,\sigma,\tau) \biggr) \, , \nonumber
\end{eqnarray}
where $\Fc(x, \sigma,\tau)$ is a generic function representing a linear combination 
of the amplitudes $A_i$, $B_i$. 
This relation can be proved by differentiating both sides with respect to $x$ and 
using the fact that the integration area vanishes at $x=1$. 
Applying relation~(\ref{eq:app_rel}) to~(\ref{eq:app_h^perp_1_sigma}) one obtains
\begin{eqnarray}
h^{\perp(1)}_1(x) \; &=& \; -\pi \int^1_x \d y \int \d \sigma \d \tau \; \Theta(\tau - y\sigma + M^2y^2) \; \biggl( \frac{\sigma-2M^2y}{2M^2} \; (-A_4) \biggr) \, , \\
\label{eq:app_d_h^perp1_1_pi}
\frac{\d}{\d x} \; h^{\perp(1)}_1(x) \; &=& \; -\pi \int \d \sigma \d \tau \; \Theta(\tau - x\sigma + M^2x^2) \; \biggl( \frac{\sigma-2M^2x}{2M^2} \; A_4 \biggr) \, .
\end{eqnarray}
Comparing now~(\ref{eq:app_h_pi}) and~(\ref{eq:app_d_h^perp1_1_pi}) it is clear that 
the LIR in~(\ref{eq:LIR4}) is fulfilled if the amplitudes $B_i$ (here $B_2$ and $B_3$) 
are absent.

%
% References
% =========
%

\end{document}